# Disparities in ridesourcing demand for mobility resilience: A multilevel analysis of neighborhood effects in Chicago, Illinois

**Elisa Borowski** [PhD Candidate]
Department of Civil and Environmental Engineering
Northwestern University, Technological Institute, 2145 Sheridan Road, Evanston, IL 60208 USA
elisaborowski2022@u.northwestern.edu

**Jason Soria** [PhD Candidate]
Department of Civil and Environmental Engineering
Northwestern University, Technological Institute, 2145 Sheridan Road, Evanston, IL 60208 USA
jason.soria@u.northwestern.edu

**Joseph Schofer** [Professor]
Department of Civil and Environmental Engineering
Northwestern University, Technological Institute, 2145 Sheridan Road, Evanston, IL 60208 USA
j-schofer@northwestern.edu

**Amanda Stathopoulos*** [Assistant Professor]
* (Corresponding author)
Department of Civil and Environmental Engineering
Northwestern University, Technological Institute, 2145 Sheridan Road, Evanston, IL 60208 USA
a-stathopoulos@northwestern.edu

**ABSTRACT**

Mobility resilience refers to the ability of individuals to complete their desired travel despite unplanned disruptions to the transportation system. The potential of new on-demand mobility options, such as ridesourcing services, to fill unpredicted gaps in mobility is an underexplored source of adaptive capacity. Additionally, whether improvements in mobility resilience provided by on-demand transportation are equitably distributed across urban areas remains unknown. Applying a natural experiment approach to newly released ridesourcing data, we examine variation in the gap-filling role of on-demand mobility during sudden shocks to a transportation system by analyzing the change in use of ridesourcing during unexpected rail transit service disruptions across the racially and economically diverse city of Chicago. By employing a multilevel mixed model, we control not only for the immediate station attributes where the disruption occurs, but also for the broader context of the community area and city quadrant in a three-level structure. Thereby the unobserved variability across neighborhoods can be associated with differences in factors such as transit ridership, or socio-economic status of residents, in addition to controlling for station level effects.

    Our findings reveal that individuals use ridesourcing as a gap-filling mechanism during rail transit disruptions, but there is strong variation across situational and locational contexts. Specifically, our results show larger increases in transit disruption responsive ridesourcing during weekdays, nonholidays, and more severe disruptions, as well as in community areas that have higher percentages of White residents and transit commuters, and on the more affluent northside of the city. These findings point to new insights with far-reaching implications on how ridesourcing complements existing transport networks by providing added capacity during disruptions but does not appear to bring equitable gap-filling benefits to low-income communities of color that typically have more limited mobility options. We explore the significance of these observed trends in adaptable mobility through the lens of mobility equity and discuss policy recommendations to address this widespread phenomenon within a rapidly evolving urban mobility context.

**Keywords:** ridesourcing, transit disruption, multilevel (random effects) mixed intercept model, natural experiment, equity, resilience

## 1. Introduction

Unexpected transit disruptions, service interruptions due to accidents, infrastructure breakdowns, and passenger distress are common occurrences in urban transit systems. The desire of riders is usually to continue their journeys to a timely completion – to be resilient in their travel. The current growth of ridesourcing services offers a novel opportunity for urban mobility resilience by filling unpredicted gaps in transit operations. In this study we examine the role of ridesourcing to enhance adaptive mobility to disruptions by complementing traditional services (Reggiani et al., 2015). Specifically, we study the variation in this type of resilience across communities.

We analyze the equity in ridesourcing for mobility resilience for the city of Chicago. Notably, Chicago is home to the second largest transit network in the U.S. with the Chicago Transit Authorities (CTA) serving 3.5 million riders (CTA, 2020a) with nearly 16 million rail rides each month (CTA, 2020b). While unequal access to essential resources is common in many U.S. cities, Chicago contends with historically rigid, spatially-defined, social and economic inequality that is frequently linked to race. For example, the income divide between White households and minority households is wider in Chicago than it is across the nation as a whole (Asante-Muhammed, 2017). Urban mobility systems typically grapple with multiple layers of inequitable mobility investments and service-access that determine service quality for different population segments. The city of Chicago is subject to urban mobility inequity both at the service level (e.g., poor mobility accessibility coverage), as well as disproportionate impacts (e.g., lack of pedestrian-friendly infrastructure or biased policing) in low-income communities (Krapp, 2020).

To understand how well a mobility system serves diverse community members, it is essential to understand the interplay between modes in the transportation system. On the one hand, public transit addresses the transportation needs of those with mobility disadvantages, implying that any disruptions in transit could disproportionately affect under-resourced communities of color. On the other hand, on-demand mobility services can offer expanded flexible mobility, although there is evidence of disparities in the use of ridesourcing related to spatial and sociodemographic diversity (Soria et al., 2020). This has led the local planning agency CMAP to highlight the need to study the potential benefits and pitfalls of new mobility technologies, such as ridesourcing, with regard to accessibility, affordable mobility, and local quality of life (CMAP, 2018).

Our awareness of existing disparities in mobility begs a fundamental question when analyzing the substitution of ridesourcing services during transit disruptions: *Is this disruption recovery equitable; that is, are under-resourced transit riders benefitting from ridesourcing-based mobility resilience on par with other travelers*? In this project, we take a natural experiment approach where we systematically identify major transit disruptions over the course of a year and match them with large scale, spatio-temporal ridesourcing trip data from the city of Chicago. We then develop a multilevel model (MLM) to examine the degree to which ridesourcing demand surges during transit disruptions, providing evidence of an adaptive ridership response. The multilevel structure is tailored to account for variation in spontaneous mobility resilience across the city and to explore whether it occurs due to neighborhood differences. To examine this potential adaptation strategy among transit riders, we compare the demand for ridesourcing during unplanned rail transit disruptions to the baseline demand while also controlling for the time of day, day of the week, and location.

The main contributions of this study are the insights it provides into: (1) whether ridesourcing is used as a gap-filling transportation mode during transit disruptions in Chicago, (2) whether its utilization for this purpose is distributed equitably across the city in terms of racial and economic circumstances, and (3) whether variation in ridesourcing demand during disruptions is attributable to station-level, community-level, or quadrant-level contexts. Our findings provide evidence of significant localized station-level effects, such as the timing of the disruption. Importantly, we also discover that the community area where the station is located is responsible for the majority of the variation in observed disruption-based ridesourcing substitution. Specifically, the racial composition and transit commuting rates show significant interaction with the aforementioned station-level factors. These insights contribute to an improved understanding of how riders cope with disruptions in different communities. In terms of practice, we discuss how our findings can guide more equitable communication strategies for transportation agencies and potential partnerships with private on-demand mobility service operators to treat mobility as a service regardless of transportation mode.

## 2. Literature review

### 2.1. Impacts of transit disruptions

Transit disruptions may be long-term or unplanned, and their impacts can cause riders to shift to traditional modes (such as cars and buses) or on-demand mobility (such as bikeshare and ridesourcing). The effects of long-term transit disruptions on transportation behavior have been widely studied over the past decade (Marsden and Docherty, 2013; Pnevmatikou et al., 2015; Pu et al., 2017; van Exel and Rietveld, 2001; Zhu et al., 2017). Long-term transit line closures may extend for several months, providing riders with time to adjust their behavior, including departure time, route, and mode, temporarily or permanently. Often during long-term disruptions, such as transit strikes, the majority of travelers switch to cars (van Exel and Rietveld, 2001; Zhu et al., 2017). This may be a function of available resources, as those who are less likely to travel by car during transit disruptions include lower income individuals, women, and individuals with more flexible work schedules (Pnevmatikou et al., 2015).

Mode-shifting behavior during long-term transit disruptions depends not only on sociodemographics but also on the surrounding context, such as the city in which it occurs. Long-term rail transit disruptions in Washington, D.C. are associated with increased bus ridership (Pu et al., 2017), especially among lower income individuals (Zhu et al., 2017). By contrast, an analysis of smart-card data for Chicago rail riders shows that the majority of riders continue to use rail during planned maintenance with a minor share shifting to bus (Mojica 2008). Furthermore, the effects of long-term transit disruptions have resulted in a permanent decline in transit ridership across Europe and the U.S. (van Exel and Rietveld, 2001; Zhu et al., 2017). Similarly, in Chicago, lengthy disruptions from track operations led to an estimated 3.9% of riders abandoning transit (Mojica, 2008).

In the few cases of research on long-term transit disruptions that consider on-demand mobility, the focus has been largely on bikeshare. Findings show temporary increases in bikesharing demand during transit strikes or maintenance projects, suggesting the ability of on-demand transportation to increase mobility resilience (Fuller et al., 2012; Kaviti et al., 2018; Pu et al., 2017; Saberi et al., 2018).

Compared to the extensive body of research on planned, long-term transit disruptions, research on *unplanned, short-term* events has been scant (Sun et al., 2016). Very recently, however, research has begun to consider the role of ridesourcing by transit users in response to unplanned service disruptions (Rahimi et al., 2020). One survey-based study, which is among the first to research transit rider behavior during unplanned service disruptions in Chicago, examined whether riders would cancel their trip, change destinations, or change modes to shuttle bus, ridesourcing, taxi, personal vehicle, or carpool (Rahimi et al., 2020). The findings reveal that the individuals who would shift to ridesourcing during a transit disruption tend to be millennials, have a higher level of education, have a smartphone, or have prior ridesourcing experience (Rahimi et al., 2020). In the survey-based study, race was not found to be a significant factor in the hypothetical shift to ridesourcing.

While earlier work has revealed the role of bikeshare as a gap-filling mechanism during long-term transit disruptions, so far, ridesourcing has only been investigated as a mode-adaptation strategy for *hypothetical* unplanned transit disruption scenarios. The current study is among the first to use a natural experiment to examine the impacts of unplanned, short-term transit disruptions on the usage of ridesourcing across the city of Chicago. We believe that examining *unplanned* transit disruptions can improve understanding of adaptation strategies that substitute on-demand mobility for fixed transportation services. Herein, we develop an MLM model to examine the magnitude of ridesourcing demand-surges due to transit disruptions. More importantly, we analyze the variation, especially related to racial composition, in the gap-filling utilization of ridesourcing and compare the disruption source and broader community-level contexts to explain this observed variation.

## 2.2. On-demand mobility usage and equity

Due to the limited access of on-demand mobility by individuals with disabilities, low income or historically marginalized communities, rural populations, under-banked households, and individuals without smartphones, the ridesourcing business model has been accused of being based on privileged access (Daus Esq., 2016). Studies show consistently that ridesourcing users are typically young, male, higher income (Zhang and Zhang, 2018), highly educated, full-time workers (Shamshiripour et al., 2020), own fewer vehicles per household, and live closer to transit stations (Deka and Fei, 2019). Similarly, users of a commute-focused vanpooling service formerly operating in Seattle were mainly White, male, highly educated, higher income, and millennials (Lewis and MacKenzie, 2017). Other works highlight contradictory findings on declared hailing-transit substitution according to income segment and service quality factors (Gehrke et al., 2018).

Recent work based on large-scale trip data and experiments has revealed more aggregate demand insights. In Chicago, greater ridesourcing usage is observed in areas with higher population and employment density, land-use diversity, household incomes, percentages of transit commuters, and percentages of zero vehicle households (Ghaffar et al., 2020). In Seattle, areas with higher percentages of racial minorities experience longer wait times for ridesourcing services at night after adjusting for differences in income (Hughes and Mackenzie, 2016), and in New York City, ridesourcing pickups are found to be less common in lower income areas (Jin et al., 2019). Additionally, evidence of racial and gender discrimination has been identified in ridesourcing practices in Boston and Seattle (Ge et al., 2016). Although a smaller body of research has found evidence of more equitable ridesourcing practices in some cities, such as longer wait times in areas with higher average income in Seattle (Hughes and Mackenzie, 2016) or more frequent ridesourcing in low-income neighborhoods in Los Angeles by one trip per month while controlling for residential location (Brown, 2019b), ridesourcing inequity is still a problem that needs to be addressed in many cities and for many groups. Furthermore, findings that racial discrimination against riders may be worse among taxis than ridesourcing (Brown, 2019a), still does not constitute evidence that ridesourcing is equitable on the whole.

In terms of bikeshare, Chicago has an embattled history of socioeconomic segregation leading to starkly different daily use patterns. Biehl et al. (2018) shows that Divvy uptake is lower in the less affluent southside even after controlling for station tenure and density. An early investigation of newly released Chicago ridesourcing data suggests a similar concentration of rides in the more affluent north and central quadrants of the city (Soria et al., 2020). This context of on-demand mobility usage and equity helps to inform our variable selection and interpretation of revealed ridesourcing substitution behavior as it relates to sociodemographics at the station, community area, and city quadrant levels. This perspective offers a valuable contribution to the literature, because equity aspects of transportation resilience have traditionally been understudied (Mattsson and Jenelius, 2015).

## 3. Data description

### 3.1. Chicago transit disruption data

Twenty-eight Chicago Transit Authority rail transit disruptions, listed in **Table 1,** are identified as having occurred during the period of November 2018 through October 2019 using a Google News search for the phrase "CTA disruption". We filtered cases to only include significant disruptions (i.e., lasting a minimum of one hour). The event-specific variables include the location (in terms of city side, quadrant, community area, and station), number of other stations impacted, disruption cause, disruption duration, deployment of shuttle buses, outside air temperature, weekday status, holiday status, peak hour status, late night status, and the cause being a medical emergency (CTA, 2019).

### 3.2. Chicago ridesourcing data

The ridesourcing data used in this study were obtained from the city of Chicago data portal (Chicago Data Portal, 2019). The transportation network companies in the dataset include Uber, Lyft, and Via. The entire dataset consists of over 152 million trips spanning the period of November 2018 through October 2019. The variables that we analyze in this study are trip start date and time, trip miles, pickup community area, and pickup census tract. The ridesourcing trip data is matched to the identified transit disruption days and comparable baseline operations. The starting location of each ridesourcing trip is identified by census tract or community area, and if that location is within a 0.25-mile radius of a disrupted transit station, the trip is included in the analysis. This frequently-used walking estimate (Younes et al., 2019; Zhao et al., 2003) is applied to account for riders who source rides on their way to or from the impacted transit station or who step away from the potential crowd surrounding the impacted station to facilitate pick-up by their ridesourcing driver. To generate a robust four-day ridesourcing demand baseline, trip counts during the disruption time period are averaged across the same day of week and time of day as the disruption for two weeks prior to the event and two weeks following. This assumes mode shifting behavior from transit to ridesourcing would typically not continue beyond two weeks following a single, unplanned transit disruption that lasts on average 2.5 hours. While this may be a conservative estimate, it avoids the risk of confounding changes in station accessibility and seasonality.

*3.3. Chicago community area data*

The city of Chicago is divided into 77 community areas, which can be further aggregated into four quadrants or sectors of the city: central, north, west, and south. The community area variables that we include in our analysis are the number of bus stations, number of Divvy stations, station ridership, population density, area, airport presence, total population, median household income, percentage of zero vehicle households, percentage of commuters taking transit, and percentage of residents who self-identify as predefined categories of Hispanic or Latino, White non-Hispanic, Black non-Hispanic, Asian non-Hispanic, or any other race and ethnicity category, according to five-year estimates from the American Community Survey (U.S. Census Bureau, 2017) and as reported by the Chicago Transit Authority (CTA, 2019), the Chicago Metropolitan Agency for Planning (CMAP, 2019), and Divvy (Divvy, 2020). Sociodemographic data on riders of transit and ridesourcing are not available, so the socioeconomic characteristics of the surrounding geographic areas are used instead. **Fig. 1(a)** depicts the locations of 28 transit disruptions, the surrounding impacted stations, community area and quadrant boundaries, and the spatial distribution of community area **(b)** median household income, **(c)** percentage of people of color, **(d)** percentage of transit commuters, **(e)** population density, **(f)** total transportation network company trips in 2019, and **(g)** percentage of active mode commuters. The maps in Fig 1 confirm the narrowly concentrated demand for ridesourcing and active mobility commuting in the central part of the city (**f** and **g**), the concentration of population and transit commuting in the northern quadrant (**e** and **d**), and the largely opposite heat-maps of income compared to the majority non-white racial breakdown (**b** and **c**).

**Table 1** Twenty-eight unplanned transit disruptions in Chicago from November 2018 through October 2019

| Number | Day | Date | Start time | End time | Quadrant | Location | Impacted span | Impacted stations | Peak hour | Shuttle bus |
|---|---|---|---|---|---|---|---|---|---|---|
| 1 | Tuesday | 11/06/18 | 5:00 | 6:00 | West | Western | Pulaski to Racine | 5 | | ✓ |
| 2 | Monday | 11/12/18 | 13:30 | 16:30 | North | Rosemont | Harlem to O'Hare | 4 | | ✓ |
| 3 | Monday | 11/26/18 | 9:00 | 12:15 | West | Cicero | 54th/Cermak to Pulaski | 2 | ✓ | |
| 4 | Thursday | 12/06/18 | 17:00 | 18:00 | Central | Jackson | Jackson | 1 | ✓ | |
| 5 | Wednesday | 12/12/18 | 5:00 | 8:30 | South | 63rd | 47th to 95th/Dan Ryan | 7 | | ✓ |
| 6 | Monday | 12/17/18 | 8:00 | 10:00 | North | Belmont | Addison to Fullerton | 5 | ✓ | ✓ |
| 7 | Saturday | 01/12/19 | 12:30 | 14:00 | South | 47th | 63rd to Sox-35th | 4 | | ✓ |
| 8 | Sunday | 01/20/19 | 9:00 | 10:30 | North | Jarvis | Belmont to Howard | 14 | | ✓ |
| 9 | Thursday | 02/14/19 | 13:00 | 16:00 | West | Clinton | Ashland to Washington/Wabash | 6 | | ✓ |
| 10 | Thursday | 03/12/19 | 21:00 | 3:00 | North | Rosemont | Jefferson Park to O'Hare | 5 | | ✓ |
| 11 | Wednesday | 04/10/19 | 19:00 | 2:00 | North | O'Hare | O'Hare to Rosemont | 2 | ✓ | ✓ |
| 12 | Wednesday | 05/01/19 | 7:20 | 8:20 | North | North/Clybourn | Cermak-Chinatown to Fullerton | 5 | ✓ | |
| 13 | Monday | 05/06/19 | 16:00 | 18:00 | North | Argyle | Argyle | 1 | ✓ | |
| 14 | Sunday | 05/12/19 | 14:00 | 16:00 | North | Bryn Mawr | Addison to Howard | 14 | | |
| 15 | Thursday | 06/06/19 | 11:00 | 16:30 | South | 47th | Ashland/63rd to Roosevelt | 10 | | ✓ |
| 16 | Monday | 06/10/19 | 9:00 | 10:00 | North | O'Hare | O'Hare to Rosemont | 2 | ✓ | |
| 17 | Wednesday | 06/12/19 | 19:20 | 20:20 | North | North/Clybourn | Cermak-Chinatown to Fullerton | 11 | ✓ | |
| 18 | Thursday | 06/20/19 | 10:15 | 13:30 | South | 35th/Archer | Halsted to Midway | 7 | | ✓ |
| 19 | Tuesday | 06/25/19 | 7:30 | 8:45 | West | Kedzie-Homan | Kedzie-Homan | 1 | ✓ | |
| 20 | Thursday | 06/27/19 | 12:30 | 15:00 | South | 69th | 63rd to 95th/Dan Ryan | 5 | | ✓ |
| 21 | Saturday | 09/07/19 | 14:00 | 15:15 | North | Belmont | Fullerton to Kimball | 15 | | |
| 22 | Tuesday | 09/24/19 | 9:00 | 10:30 | North | Sedgwick | Sedgwick | 1 | ✓ | |
| 23 | Thursday | 09/26/19 | 1:00 | 4:00 | North | Rosemont | Harlem to O'Hare | 4 | | |
| 24 | Thursday | 09/26/19 | 17:45 | 22:00 | North | Jarvis | Belmont to Howard | 14 | ✓ | |
| 25 | Saturday | 10/05/19 | 22:45 | 2:15 | North | Granville | Belmont to Howard | 14 | | |
| 26 | Tuesday | 10/08/19 | 15:15 | 16:15 | South | 63rd | Roosevelt to 95th/Dan Ryan | 10 | | ✓ |
| 27 | Wednesday | 10/30/19 | 15:00 | 16:00 | North | Howard | Belmont to Howard | 14 | | |
| 28 | Thursday | 10/31/19 | 16:15 | 18:30 | Central | Harrison | Cermak-Chinatown to Fullerton | 11 | | ✓ |

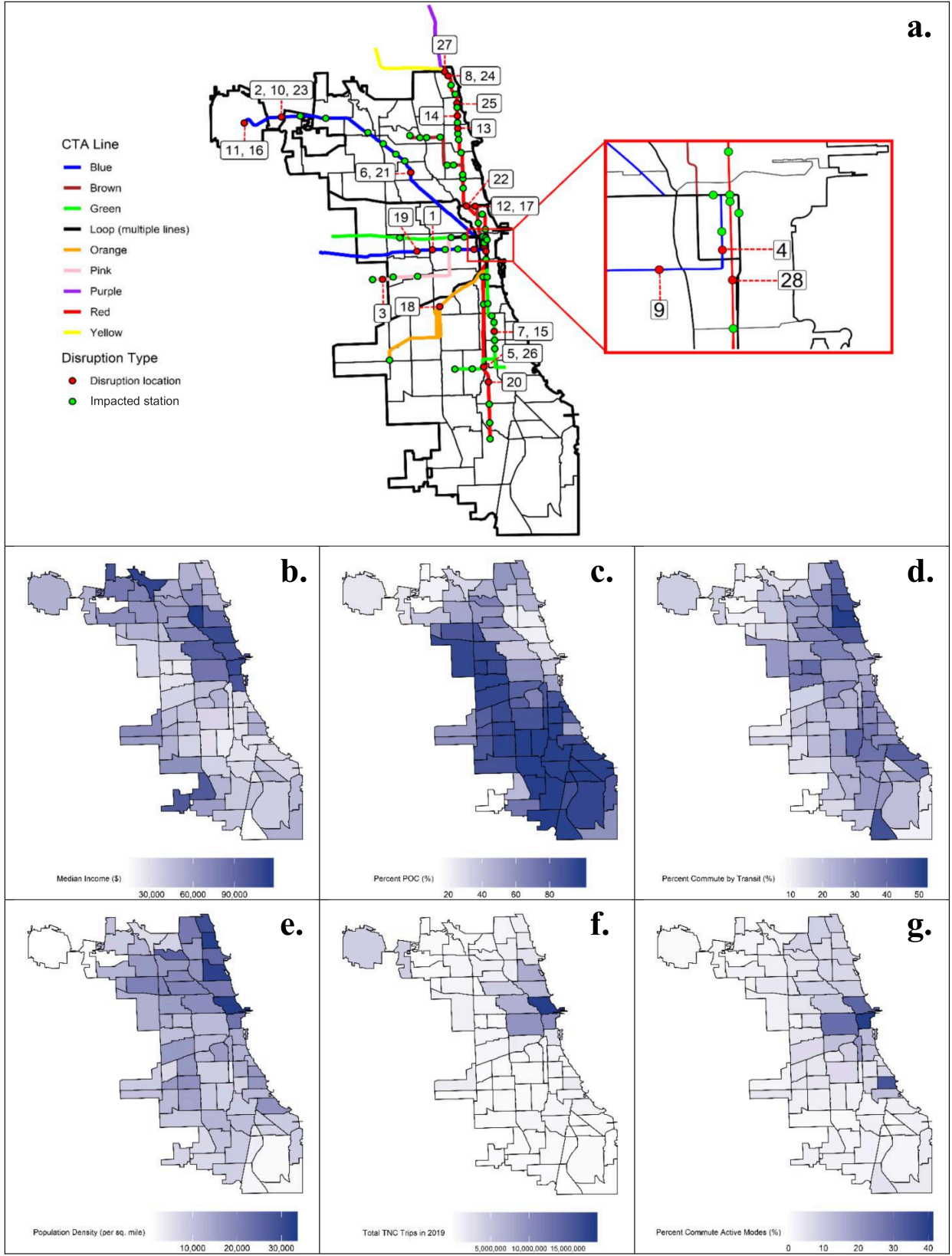

**Fig. 1. (a)** Map of Chicago showing rail transit lines and disruption events in relation to community area and quadrant boundaries. *Insert represents the Loop central business district*. Heatmaps indicate the spatial distribution of **(b)** median household income, **(c)** percent people of color, **(d)** percent commute by transit, **(e)** population density, **(f)** total transportation network company ridesourcing trips in 2019, and **(g)** percent commute by active modes.

## 4. Modeling methodology

*4.1. Multilevel mixed model theory*

Multilevel mixed (MLM) modeling is designed to properly account for the hierarchical nesting of data and effects happening at different levels (Goldstein, 2003; Julian, 2001; Wampold and Serlin, 2000). MLM models provide a mechanism for analyzing datasets where observations (in this case, station disruptions) are nested within higher-order spatial contexts, such as neighborhoods. In the past, MLM or hierarchical models have been used to represent the structure of social relations within personal networks (Carrasco and Miller, 2009), temporal changes in bike share trips (El-Assi et al., 2017), and transit demand between origin-destination station pairs (Iseki et al., 2018). We use a multilevel regression analysis to identify the station level, community area level, and city quadrant level factors associated with systematic variations in ridesourcing demand during transit disruptions. We can thereby examine explanatory variables at each level of the data hierarchy, and in doing so, control for community area effects on station ridership variation.

The advantage of using the multilevel structure is the ability to estimate the variability in results that can be attributed to neighborhood (e.g., community area) effects rather than only to individual station effects. By carefully controlling variable-inclusion at the appropriate level, the model takes into account correlations between observations within the same group (i.e., a given community area) as distinct from correlations between groups (Jones and Duncan, 1996). Instead, a standard one-level regression model would ignore group-level distinctions (for example, different commuting patterns in different communities) and group level correlations (for example, similar patterns of use among stations in the same community related to the income-level of riders). A useful way to think of MLM models is as a structure sitting between two modeling extremes when groupings are known: fully pooled and fully unpooled specification (Gelman and Hill, 2007). A fully pooled model treats group-level variables as individual variables, thereby ignoring group-level distinctions. The opposite extreme, a fully unpooled model, asserts that the groups are so completely different that they cannot be associated in the same model. The MLM model offers a compromise between complete distinction of groups and the complete ignorance of group-level effects by modeling individual-level fixed effects as well as distributional assumptions on the random effects.

To address the research question of ridesourcing surges triggered by transit disruptions, we control not only for the immediate station attributes where the disruption occurs (Level 1), but also for community area (Level 2) and quadrant (Level 3) factors in a three-level structure. **Fig. 2** shows the hierarchical, three-level model framework. The dependent variable is the number of ridesourcing trips compared to the baseline demand two weeks prior and two weeks following the disruption (i.e., individual station observations). Covariates related to the disruption cause, context, and timing are included as explanatory variables at this level, in line with Mojica (2008) and Pu et al. (2017). We further investigate whether the fact that stations are nested within community areas and major quadrants plays a role in ridesourcing demand shifts. It is likely that a comparable disruption can generate different mode-shifting effects depending on where it is located, owing to the different composition of travelers and availability of alternative modes. Specifically, the broader context is controlled for by including sociodemographic and mobility factors measured at the community level that are in turn aggregated to the quadrant level of analysis. It is worth noting that since the disruptions we measure result from a natural experiment, we are unable to control exhaustively for all combinations of factors that are at play within and between community areas. Therefore, we include random intercept effects at each of the lower-nested group levels to partition the unexplained variability effects on the dependent variable.

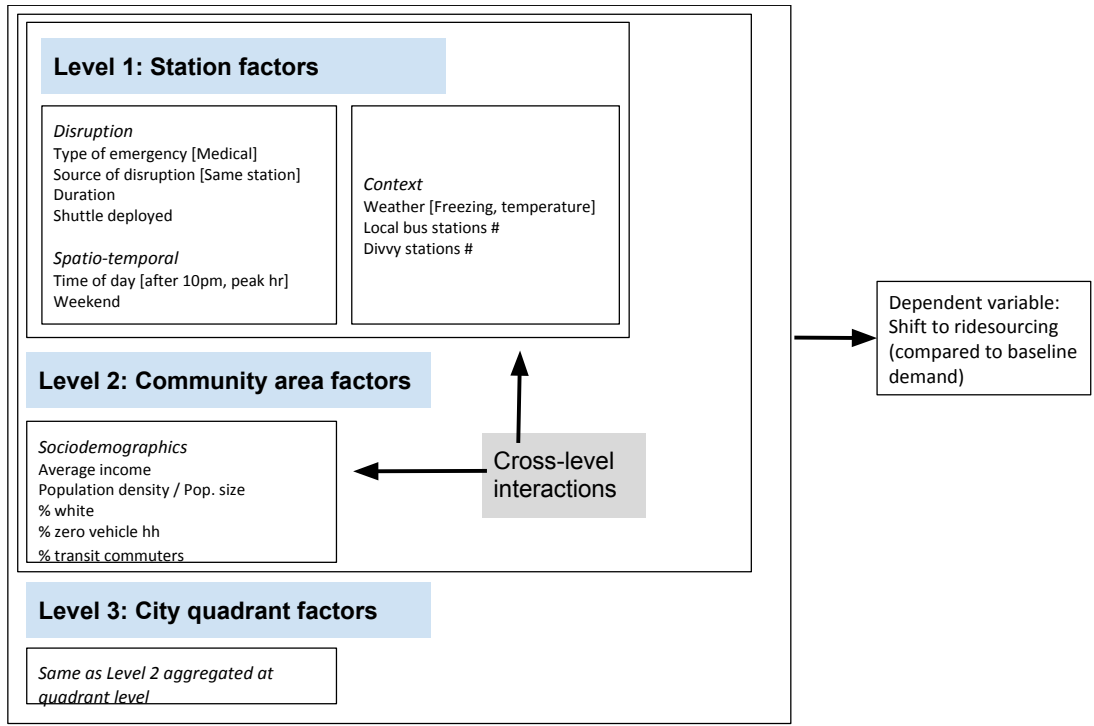

**Fig. 2.** Framework for the multilevel mixed model of ridesourcing shifts caused by transit disruptions. Station disruptions are nested within community areas, which in turn are nested within quadrants. *Considered variables are listed for each level.*

Conceptually, the model can be articulated as regression equations occurring at different levels where each group-level coefficient has its own regression equation. Following Gill and Womack (2013), the general three-level structure is defined in **Eq. 1** as:

$$y_{ijk} = \beta_{0jk} + \beta_{1jk} x_{1ijk} + \varepsilon_{ijk} \qquad (1)$$

where *i* represents the station, *j* represents the community area, and *k* represents the quadrant. $\beta_{0jk}$ is the (random) intercept measuring average ridesourcing use (defined in **Eq. 2**), and $x_{1ijk}$ is a predictor, such as the average daily transit use measured at the station level, while $\beta_{1jk}$ is the (random) slope depicting the relationship between the station-variables and the change in ridesourcing demand (as defined in **Eq. 3**). The error term $\varepsilon_{ijk}$ relates to station-level effects.

*Borowski et al.* 11By including Level 2 and 3 explanatory variables in the model, we uncover context-level effects. Namely, we account for variability in coefficients at the station level owing to community area or city quadrant level factors. Level 2 includes variables aggregated to the community area level expected to impact all stations in the area. This can be thought of as being equivalent to the way in which student educational performance is affected by their classroom teacher in a way that is distinct from the effects of their individual factors or from more aggregate school-level effects. The subscript *jk* denotes the distinct community area impacts. The ɣ has numbered subscripts, the first denotes the intercept (0) or slope (1), while the second subscript denotes the independent variable. At Level 2, the general regression equations are defined as:

$$\beta_{0jk} = \gamma_{00k} + \gamma_{01k} x_{2jk} + u_{0jk} \tag{2}$$
$$\beta_{1jk} = \gamma_{10k} + \gamma_{11k} x_{2jk} + u_{1jk} \tag{3}$$

Where the random intercept $\beta_{0jk}$ is a function of $\gamma_{00k}$, the grand mean of ridesourcing demand surges across stations in the community (defined below in **Eq. 4**). Departures from this average intercept represented by $x_{2jk}$ are community-level predictors with $\gamma_{01k}$ denoting the random slope for community level predictors (**Eq. 5**), and $u_{0jk}$ is the unique effect associated with communities assumed to have a multivariate normal distribution. The random slope $\beta_{1jk}$ is a function of $\gamma_{10k}$ representing the average effect of the station-level predictors (i.e. the slope over all stations shown in (**Eq. 6**)). Departures from the slope (i.e., random effects) over station predictors are represented by the $\gamma_{11k}$ coefficient (**Eq. 7**) that would be removed for a random intercept-only model (as in the current case).

At Level 3, variables vary by quadrant and apply to all individual cases and community areas assigned to this group. Therefore, they contain the subscript *k* as opposed to *ijk* or *jk*. At Level 3, the separate regression equations for the intercepts and slopes are defined as:

$$\gamma_{00k} = \delta_0 + \delta_4 x_{3k} + u_{00k} \tag{4}$$
$$\gamma_{01k} = \delta_2 + \delta_5 x_{3k} + u_{01k} \tag{5}$$

$$\gamma_{10k} = \delta_1 + \delta_6 x_{3k} + u_{10k} \tag{6}$$
$$\gamma_{11k} = \delta_3 + \delta_7 x_{3k} + u_{11k} \tag{7}$$

where $\delta_0$ is the intercept shared by all individual cases, $\delta_1, \delta_2$, and $\delta_3$ are the main effects, $\delta_4, \delta_5$, and $\delta_6$ are two-way interactions, and $\delta_7$ is a three-way interaction.

In our specific modeling, the outcome variable of the three-level hierarchy $y_{ijk}$ is defined as the change in ridership over baseline. After specification testing, the final model takes the forms shown in **Eq. 8-16**. The model includes a random intercept $\beta_{0jk}$ and two main effects ($non\_holiday_{ijk}$ and $peak\_hour_{ijk}$) at Level 1, shown in **Eq. 8**. Level 2 brings in contextual variables used to explain variability in ridesourcing demand via cross-level interactions. That is, now we model the intercept and slopes explicitly, and include level-one and level-two independent variables interacted to describe variation in the intercept. **Eq. 9-11** shows the random intercept $\gamma_{00k}$ and the cross-level interaction terms ($percent\_white_{jk} \times peak\_hour_{ijk}$ and $percent\_transit_{jk} \times disruption\_source_{ijk}$). Level 2 also specifies $\beta_{1jk}$ and $\beta_{2jk}$ which represent the parameter slopes with $\gamma_{10k}$ and $\gamma_{20k}$. Level 3 includes the random intercept $\delta_0$ and one quadrant level interaction ($north\_quad_k \times shuttle_{ijk}$) that is found to generate variability in ridesourcing (eq. **12**), with remaining parameters $\delta_1$ and $\delta_2$ denoting the fixed slope coefficients. The disturbance parameters are included at the community $u_{0jk}$ and quadrant levels $u_{00k}$ (**Eq. 15-15**). It is important to note that the cross-level interactions explain a significant amount of variance of ridesourcing demand changes in addition to that already explained by the station-level equations.

$$\text{Level 1 model}$$
$$y_{ijk} = \beta_{0jk} + \beta_{01k,non\_holiday}\, non\_holiday_{ijk} + \beta_{peak\_hour} peak\_hour_{ijk}\, peak\_hour_{ijk} + \varepsilon_{ijk} \tag{8}$$

$$\text{Level 2 model: Random Intercept \& Cross Level interactions}$$
$$\beta_{0jk} = \gamma_{00k} + \gamma_{01k}\, x_{percent\ white, jk} \times x_{peak-hour,\ ijk}$$
$$+ \gamma_{02}, x_{percent_{transit}, jk} \times x_{disruption\_source_{ijk}} + u_{0jk} \tag{9}$$
$$\beta_{1jk} = \gamma_{10k} \tag{10}$$
$$\beta_{2jk} = \gamma_{20k} \tag{11}$$

$$\text{Level 3 model: Random Intercept \& Cross Level interaction}$$
$$\gamma_{00k} = \delta_0 + \delta_{NorthShuttle}\, north\_quad_k \times shuttle_{ijk} + u_{00k} \tag{12}$$
$$\gamma_{10k} = \delta_1 \tag{13}$$
$$\gamma_{20k} = \delta_2 \tag{14}$$
$$u_{0jk} \sim \mathcal{N}(0, \sigma_d^2) \tag{15}$$
$$u_{00k} \sim \mathcal{N}(0, \sigma_e^2) \tag{16}$$

## 5. Results

### 5.1. Descriptive results

#### 5.1.1. Neighborhood differences in ridesourcing demand-surge during disruptions

The descriptive analysis suggests that a significant surge in the use of ridesourcing does occur following no-notice transit line disruptions, as suggested in **Fig. 3(a)**. This depicts a high-impact northside case in Lake View at the Belmont station (the source of the disruption) on a Monday in December during morning peak hours. The disruption cause was a train striking a person. For reference, the baseline ridesourcing demand for this timespan and location is 807 rides. The disruption is associated with a significant surge in ridesourcing requests, totaling 2,883 and corresponding to an increase of 257% (a t-test of the disruption versus the baseline means has a p-value of 0.00001). However, ridesourcing adaptation is not uniform across the city. A different case is shown in **Fig. 3(b)** of a similar disruption event, this time occurring in an under-resourced westside neighborhood where there appears to be limited shifting towards on-demand services. This shows a low-impact case in East Garfield Park at the Kedzie station (the source of the disruption). Similar to the Belmont disruption, it occurred during weekday morning peak hours and was caused by a person on the tracks. The baseline ridesourcing demand for this time and location is a fraction of that at Belmont: 89 rides. The number of ridesourcing rides during the disruption event is lower than the baseline at 76 (-15%), which is not a significant change (a t-test of the disruption versus the baseline means has a p-value of 0.3984). Given that Lake View has a median household income of $86,119 and 79% of its residents are White, while East Garfield Park has a median household income of $23,116 and 5.6% of its residents are White, this behavioral difference could be tied to racial and economic inequity. To examine the different patterns of ridesourcing demand shifts triggered by transit disruptions systematically, we turn to the model results.

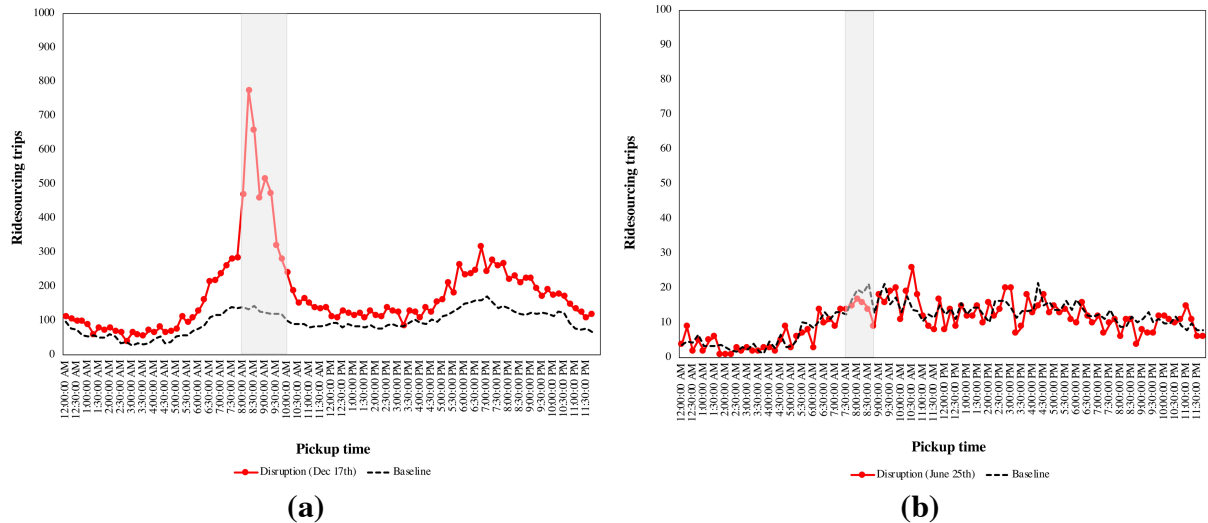

**Fig. 3.** Ridesourcing demand during disruptions (solid red line) compared to baseline (dotted black line) at: **(a)** the Belmont station in the north quadrant and **(b)** the Kedzie station in the west quadrant. *Disruption duration marked in gray shading. The y-axes are scaled to reflect the different levels of baseline demand (10:1).*

*5.2. Multilevel mixed modeling results*

As shown in **Table 2**, the MLM model includes both station-level (Level 1) fixed effects (*non-holiday disruption* and *peak hours disruption*), which are similar to standard regression parameters, and explanatory features that reflect the context surrounding the station: namely, two cross-level random effects (*percent White during peak hours* and *percent transit commuters at the source of disruption*), and one quadrant-level effect (*shuttle deployment in the north quadrant*). All parameters are significant to a 98.9% level of confidence or greater except for the model constants. The model also includes random intercepts for the community area (Level 2) and city quadrant (Level 3) effects.

First, we estimate a null model that partitions the variance at each level without including any explanatory variables. This enables us to calculate the intraclass correlation (ICC), also known as the variance partition coefficient, for three levels, following Snijders and Boskers (1999). This statistic measures the correlation among data at the lower levels to determine how much of the variation in ridesourcing demand is accounted for by variation among community and quadrant level factors. A *smaller* variance partition coefficient indicates that the variation in ridesourcing shifts is attributable more to variation among *lower-level* units (such as stations) than to that among upper-level units (like community areas and quadrants). The empty MLM model (Model 1) thereby provides an estimate of baseline variance of ridesourcing demand shifts due to factors beyond the immediate station. The intra-quadrant correlation is small and suggests that 20% of the variance of the dependent variable is due to quadrant-level effects. In contrast, the intra-community correlation reveals that the largest share of total variance (43%) is related to community-level factors, suggesting that the lowest level of analysis (the station level) explains the remaining variance (37%). This indicates the largest share of the variation in ridesourcing demand substitution is related to factors that occur across communities, followed by the station level.

**Table 2** Multilevel mixed modeling results

|  | Model 1 | | | Model 2 | | | Model 3 | | |
|---|---|---|---|---|---|---|---|---|---|
| **Fixed part** | Coef. | z value | P > |z| | Coef. | z value | P > |z| | Coef. | z value | P > |z| |
| Non-holiday |  |  |  | 541 | 2.97 | 0.003 | 599 | 3.58 | 0.000 |
| Peak hour |  |  |  | 408 | 4.11 | 0.000 | 353 | 4.22 | 0.000 |
| Constant | -53.2 | -0.39 | 0.700 | 54.1 | 0.97 | 0.332 | 91.8 | 1.53 | 0.125 |
| *Cross-level interactions* |  |  |  |  |  |  |  |  |  |
| Peak hour * Percent white |  |  |  |  |  |  | 12.3 | 2.69 | 0.007 |
| Disruption source * Percent transit commuters |  |  |  |  |  |  | 18.4 | 2.54 | 0.011 |
| *Quadrant-level interactions* |  |  |  |  |  |  |  |  |  |
| North quadrant * Shuttle |  |  |  |  |  |  | 321 | 3.66 | 0.000 |
| **Random part** | Estimate | Std. Dev. | Std. Err. | Estimate | Std. Dev. | Std. Err. | Estimate | Std. Dev. | Std. Err. |
| Level 3: Quadrant | 55178.9 | 234.902 | 62705 | 4.92E-10 | 2.2E-05 | 9.51E-09 | 3.97E-17 | 6.3E-09 | 1.10E-15 |
| Level 2: Community area | 121121 | 348.024 | 49597 | 101766 | 319.008 | 41407.7 | 135958 | 368.724 | 43158.8 |
| Residual | 102351 | 319.923 | 20828 | 81116.4 | 284.809 | 17049.7 | 50515.2 | 224.756 | 11244.2 |
| **Fit statistics** |  |  |  |  |  |  |  |  |  |
| Log likelihood |  | -836.99148 |  |  | -822.09863 |  |  | -809.57185 |  |
| LR test |  | 12.17 |  |  | 7.15 |  |  | 17.1 |  |
| Prob > chi2 |  | 0.0023 |  |  | 0.028 |  |  | 0.0002 |  |
| AIC |  | 1681.983 |  |  | 1672.508 |  |  | 1637.144 |  |
| R2 |  |  |  |  | 0.34 |  |  | 0.33 |  |
| *Intraclass correlation* |  |  |  |  |  |  |  |  |  |
| Level 3: Quadrant |  | 20% |  |  | 0% |  |  | 0% |  |
| Level 2: Community area |  | 43% |  |  | 56% |  |  | 73% |  |

The evolution of the variance estimation deserves attention. When adding station-level independent variables in Model 2, the quadrant random intercept and thereby ICC is shown to be insignificant, while the variance is now partitioned between the station (44%) and community area levels (56%). Along the same lines, when adding cross-level effects by including variables measured at the community area level, the variance explained clearly pivots toward the community area variables (Model 3). We note that despite the Level 3 quadrant random intercept collapsing to zero, removing this variance component from the analysis causes a significant reduction in overall model fit. This analysis of random intercepts points to two important observations. First, the variance partitioning shows the importance of controlling for variables measured at the neighborhood factors that would have been overlooked in a standard regression solely focused on station substitution analysis. Moreover, the inclusion of more explanatory factors leads to the absorption of more variability in the ridesourcing demand shift at the neighborhood levels.

The main takeaway from the variance controls is the following: the differences between factors occurring across different community areas in the city are the most decisive in shaping the ridesourcing demand shifts following disruptions. We interpret this to mean that there are significant latent neighborhood effects at play in the shift to ridesourcing during transit disruptions.

*5.2.1. Station level analysis*

To model systematic variables, we follow the block entry approach consisting of the gradual addition of covariates level by level (Cohen et al., 1983), following the plan previously outlined in **Fig. 2**. First, each of the hypothesized predictors measured at the station-level are tested independently, then jointly. Owing to high variable collinearity, only two fixed-effect explanatory variables related to the timing of the disruption and a constant are included in the resulting Model 2. These statistically significant effects result in a significant improvement in model fit as measured by the deviance difference (836.99 - 822.10 = 14.89, exceeding the critical $\chi^2$ of 5.99 with alpha set at .05) and AIC reduction.

Looking at the constant, the model suggests a moderate average increase of 54 ridesourcing trips (or 15.6%) during a transit disruption, compared to baseline. To contextualize this finding, we note that the average baseline ridesourcing ridership is 347 trips across the Chicago community areas covered in the disruption analysis. This value can be seen as the ridesourcing demand that would be occurring for the same station and timespan without the disruption. With this baseline in mind, the timing of disruptions is revealed to be highly impactful. On average, when a disruption occurs on a weekday (excluding holidays), ridesourcing rides increase by 541 from baseline (a 156% increase). When a transit disruption occurs during peak hours, ridesourcing demand increases by 408 rides from baseline (a 118% surge).

The analysis suggests the existence of some citywide trends related to the timing of no-notice disruptions, likely related to less flexible trips during peak hours and weekdays. This finding is not surprising considering that business and commuting trips are more likely to be shifted to another mode than canceled, which has been shown for pre-planned disruptions (Van Exel and Rietveld, 2009) and for unreliable metro services (Pnevmatikou and Karlaftis, 2011). The novelty of our findings refers to the degree of shifting towards *on-demand ridesourcing*, a mode which has not been considered in previous work, which has been dominated by car substitution and transit replacement analysis.

For completeness, following the **Fig. 2** modeling framework, we note that we are able to find no consistently significant impact of temperature, deployment of shuttle buses, number of nearby bus or Divvy stations, nor general transit commute ridership. This is somewhat surprising given that the research on consequences of long-term transit disruption shows that the context (e.g., spatial or temporal) results in different rider adaptation strategies. Therefore, we had expected specifically that modal alternatives (e.g., buses, bikeshare, etc.) would impact ridesourcing. For example, given previous research findings on the role of bikeshare as an adaptive strategy during long-term transit disruptions in Washington, D.C. and London, we had expected to find that the availability of nearby Divvy bikeshare stations would significantly decrease the observed shift to ridesourcing during unplanned transit disruptions. However, this was not the case, possibly due in part to trip purpose, travel distance, local bikesharing culture, and/or the inconvenience of requesting a membership in response to a single unplanned disruption. Nevertheless, from the ICC calculation in the current study, we know that the community area context is the main source of unexplained systematic differences in ridership shifting strategies. These neighborhood effects likely vary as a function of ridership culture (including car, transit, and ridesourcing culture), socioeconomic and political factors, and transportation agency strategies. While it is possible that some of these effects could be measured narrowly around the disruption events (i.e., stations), we note that the MLM enables us to analyze important factors like population density, median household income, and racial composition measured at more aggregated levels. The next section seeks to pinpoint the systematic factors that can explain the observed stochasticity.

*5.2.2. Community context effects*

A fundamental goal of this multilevel analysis is to estimate the variability in ridesourcing use that can be attributed to community area level characteristics rather than to individual station factors and to identify how these components of variation change with the inclusion of predictors that quantify the context. In Model 3, we hypothesize that the context surrounding the disruption also plays a role in determining the transfer of ridership from transit to ridesourcing during no-notice events. Specifically, we hypothesize that the rate of ridesourcing substitution is associated with sociodemographic privilege, in line with earlier research (Deka and Fei, 2019). This guides the inclusion of a number of predictor variables measured at the community area (not station) level in the form of cross-level interactions, including factors such as *zero car households* and *population density*. We apply group mean centering for community area variables (Enders and Tofighi, 2007) to facilitate the interpretation of the cross-level interactions.

Model 3 reveals a significant impact of two community area level factors; *racial composition* and *percent of transit commuters*. The addition of these cross-level factors leads to significant improvements in goodness-of-fit measured by the deviance difference and AIC. While the cross-level effects look comparatively small, they need to be related to the percentage unit of the variable measurements. The positive effect associated with the interaction term for *percentage of White residents* in the community area with the dummy variable for *peak-hour travel* (a coefficient of 12.3 additional trips) provides insight into the combined effect of race composition in the local area on the previous peak-hour effect findings. Namely, across community areas, the peak-hour impact (353 added trips) is further boosted when disruptions occur in communities with higher shares of White residents. The implied difference is that, other things equal, a disruption occurring in a community area with a 10% higher share of White residents would result in a boost of 120 (or 34.6%) ridesourcing trips compared to the average peak-hour baseline. Recalling that communities of color in Chicago are more likely to be under-resourced in relation to job accessibility, transit supply, and on-demand mobility, we believe this finding is more likely a reflection of gaps in *access* to resources in areas with lower shares of White residents than of a lower *willingness* to use ridesourcing during disruptions. This finding adds to existing evidence that ridesourcing, in this case as a disruption gap-filling resource, gives more benefit to privileged user groups (Zhang and Zhang, 2018).

Additionally, a novel effect is found related to the *proportion of transit commuters* in the community area and the *incident location*. On the whole, every percentage unit increase in transit commuting in the community area results in 18 added ridesourcing trips (or a 5.2% increase). However, this effect only occurs at the station where the incident responsible for the disruption is occurring. We speculate that transit commuters more readily shift to hailing services when they experience the disruption and receive information about it *directly*. In other words, riders at the source of the disruption are likely to have more information regarding the nature of the disruptive event from official sources and other riders, which will likely factor into their adaptation strategy. On the other hand, in areas with less transit commuting, there is presumably less collective experience with disruptions and therefore a higher likelihood of shifting to other private modes.

Despite there being no unexplained systematic differences related to the quadrant level beyond Model 1, we conduct a model search to look for further impactful interactions including quadrant dummies for the four parts of the city. The model suggests a surprising finding. In the *north quadrant*, when a *shuttle is deployed*, ridesourcing rides increase by 321 instances (or 92.5%) from baseline. The deployment of replacement bus services during rail disruptions is the most common agency response (Pender et al., 2013). Yet, there appears to be an unintended negative effect of this strategy. That is, the deployment of added transit bus capacity to assist riders should not *boost* ridesourcing requests. We interpret the unexpected increase in ridesourcing to be related to the signaling effect of this action. Riders could perceive bus deployment as a strong cue for the severity of the disruption, or they may fear excessive crowding on the buses, both of which justify the decision to use ridesourcing. The north quadrant is home to the biggest share of disruptions in our dataset (16/28 or 57%), as well as heavy transit demand by commuters (**Fig 1.d**). The fact that replacement bus deployments trigger more ridesourcing substitution in the north quadrant is likely related to the higher income levels among commuters in this corridor.

*5.3. Summary of findings*

Given the situational and locational contexts impacting the use of ridesourcing as an adaptive transportation strategy, it is important to consider the potential mobility resilience inequity across the city. If ridesourcing as a gap-filling mechanism is mainly associated with mandatory travel, disruptions of greater severity (i.e., requiring the deployment of a shuttle bus), and community areas with higher percentages of White residents and transit commuters, its role in mobility resilience is specific, selective, and unlikely to address existing issues of mobility inequity. Specifically, our findings show that riders in areas with a higher proportion of non-White residents are less likely to divert to ridesourcing in times of disruption. This result does not in itself suggest supply inequity. However, when taking into account the fact that gaps in mobility services disproportionately impact communities that lack good access to transportation, the finding raises an important issue. Under-resourced communities with poorer accessibility options would benefit the most from access to more adaptability options in the face of transit disruptions, and ridesourcing could play a greater role in this adaptation portfolio. Taken together, our findings point to an opportunity to increase equitable mobility resilience by addressing the barriers that limit access to on-demand transportation. This highlights the need to consider socioeconomic constraints in disruption response planning and to fill service gaps through collaboration between transit providers and ridesourcing companies.

**6. Policy and research discussion**

*6.1. Mobility partnerships and communication*

When a major transit disruption occurs, riders may not be well-informed of the reason for the disruption or the expected duration. Typically, some information is provided to passengers already on a transit platform, and only a brief description of a service alert may be offered online. However, this information may not be timely or may be generic or incomplete, making it difficult for passengers to develop informed response strategies. In this paper, we show how riders spontaneously respond to disruptions by substituting transit with ridesourcing. In some settings this occurs despite agency efforts to appease their passengers. For example, ridesourcing demand surges *more* when shuttles are deployed or when the disruption occurs at the traveler's station.

Through improved communication, transit providers could advise travelers of when and how to seek alternative transportation means and efficiently inform ridesourcing companies of disruptions. Given this information, ridesourcing companies and drivers could work in tandem with route-around bus services to meet spikes in demand and avoid exploiting the situation through surge pricing. The flexibility of ridesourcing services offers on-call availability to provide extra capacity, while buses are able to maintain fixed routes for prolonged periods of time. In some cases, it might actually be cheaper for transit carriers to subsidized shared ridesourcing instead of delivering shuttle buses. By communicating the nature of the disruption and anticipated needs, transit agencies could engage ridesourcing drivers in adaptive, gap-filling services to address unplanned disruptions and reduce the adverse impacts experienced by transit riders.

While collaborations between transit providers and ridesourcing companies may provide a way to decrease disruption response time and assist a greater number of affected travelers, these partnerships are not without challenges. Most notably, transit providers are required to ensure fair service to all individuals in accordance with Title VI and the Americans with Disabilities Act, while ridesourcing services are not currently held to the same standards. Another crucial challenge is the negotiation of data and information sharing among public transit agencies and private transportation providers. Public-private data-sharing partnerships are an important collaborative mechanism through which governmental agencies and ridesourcing companies may develop data-supported policies to fill transportation gaps while protecting user privacy (Cohen and Shaheen, 2018).

*6.2. Adaptable mobility equity*

Our findings suggest that transit riders in areas with a higher proportion of non-White residents and in less affluent quadrants may be left out of the option to shift to on-demand ridesourcing during unplanned transit disruptions. In this section we discuss how the joint goal of providing mobility adaptiveness and achieving transportation equity, can be achieved across three levels of analysis: spatial, economic, and social.

First, existing spatial inequities in Chicago have resulted in low-income communities and communities of color experiencing long-running transportation challenges. Residents in these communities are subject to spatial mismatch, have less accessibility to public transit within walking distance, and experience longer commutes, summarized as having fewer livability opportunities (Ferrel et al., 2016). Underlying mobility inequity at the spatial level can be addressed by transportation policymakers through partnerships between public transit agencies and ridesourcing companies and by providing more micro-transit in lower-density areas.

Second, economic inequity is apparent in the distribution of resources that are needed to use ridesourcing, including direct costs (e.g., proportion of income spent on fares) and indirect costs (e.g., smartphones, credit cards, and Internet access). As urban areas become more digitally integrated, offering residents greater mobility resilience to disruptions through on-demand ridesourcing services, those on the underserved side of the digital mobility divide are left further behind. Practical measures to combat direct and indirect costs include providing lower income commuters or travelers with subsidies or vouchers for lower cost access to ridesourcing during transit disruptions, access to smartphones, multi-modal hubs with Internet access, and unbanked payment options similar to transit smartcards.

Third, social inequities refer to existing racial, cultural, and language barriers to ridesourcing usage. For example, there is a potential supply equity challenge in light of the evidence that communities of color experience longer ridesourcing wait times than non-White communities in Seattle (Hughes and Mackenzie, 2016), also suggested for Chicago (CNT, 2019). Longer wait times can be especially troublesome for peak period commuters. To address social inequities, such as supply inequity wherein the opportunity to use ridesourcing is not readily available to all communities, there is a need to consider ridesourcing incentives, safety initiatives, and driver training on socially equitable practices to minimize sociodemographic profiling. To combat poorly targeted marketing, transportation agencies could offer information campaigns and outreach programs geared toward historically underserved communities that highlight equitable transportation initiatives to improve adaptability and resilience during travel disruptions. In particular, it is important that under-resourced communities have a say in the decisions that impact them through bottom-up concept generation and participatory policymaking.

*6.3. Mobility-as-a-Service Future?*

Ultimately, agencies may shift toward the idea of Mobility as a Service. If transit operators, like the CTA, were to adopt a mandate to deliver stop-to-stop service despite unplanned disruptions, they might internalize the responsibility of providing transportation alternatives when a disruption occurs. This lack of separation between agencies and domains may represent the future of transportation operations, and public-private partnerships between transit agencies and ridesourcing companies may provide a first step toward making this future a reality. To meet service quality constraints, the on-demand rapidity of ridesourcing service response could be advantageous. If deployed ridesourcing was moreover provided as a pooled service, it might be more cost efficient for transit agencies than a shuttle service, which requires equipment maintenance by the CTA and short-notice driver availability (Shared-Use Mobility Center, 2020). Furthermore, such partnerships would be beneficial to have in place ahead of long-term transit service disruptions, such as those observed during the COVID-19 pandemic. For example, the LA Metro in the Los Angeles region was able to leverage a preexisting partnership with the ridesourcing company Via during the pandemic by expanding their role from providing first- and last-mile services to private, point-to-point trips to accommodate essential travel (Grossman, 2020). This exemplifies the ability of public-private partnerships to increase mobility resilience to unexpected disruptions.

**7. Limitations and future directions**

While this study is among the first to examine impacts of unplanned transit disruptions on the usage of on-demand ridesourcing services across the city of Chicago, some caveats warrant discussion. First, information on transit disruption and shuttle bus deployment were gathered from local news sources and are subject to source accuracy. For the analysis, assumptions were made that disruptions lasted the same duration at each affected station and that shuttle services (when provided) were deployed to all affected stations. Second, determining the spatial band where riders change travel behavior is complicated to determine and depends in part on their location when informed of the disruption, as well as their intended destination. We consider mode-switching behavior within a 0.25-mile radius of each affected station which is likely to underestimate the true degree of shifting. Particularly, mode-shifting behaviors may have occurred across a broader time-space domain, including travelers who learned of the disruption prior to departure. Third, we use aggregated measurements of community variables as a proxy for individual-level attributes, which masks variation in socioeconomic characteristics of riders. The characteristics of the actual riders aboard the train and the areas in which they reside are unknown.

Despite these limitations, which are all associated with the use of a natural experiment approach, our research contributes new insights that would be difficult to gage using smaller scale stated data. Namely, we capture the actual circumstances of the disruptions leading to adaptive use of ridesourcing. Importantly, the findings of this study bring to light which community groups are able to shift to on-demand mobility during a disruption and who is left behind to seek out other alternatives.

Our findings suggest two main avenues for future research. First, to better gage socioeconomic distributions of mobility resilience, further collaborative research should aim for a more nuanced analysis of the behavioral adaptations practiced by transit riders who do and do not shift to on-demand mobility. For example, access to disaggregated data on ridesourcing pickup locations with rider sociodemographics matched with spatio-temporal bus ridership would reveal more detailed insights into a *user*-based (in)equitable distribution of multi-modal, context-specific adaptive strategies that are enacted to complete disrupted travel (i.e., our measure of mobility resilience) across disaggregated marginalized populations. Second, the latent area effects observed in this multilevel analysis suggest that future work should investigate the impacts of mobility culture, values, and attitudes surrounding the use of ridesourcing as an adaptive mobility strategy. Better understanding of *latent* constraints would enable the tailoring of transportation policies by area to improve equity and adaptability, thereby enhancing mobility resilience. While research on the provision of information to riders and their observable reactions is extensive (Leng and Corman, 2020; Sarker et al., 2019, Ben-Elia et al., 2013, Mahmassani & Liu 1999), there is still a need for an empirical understanding of adaptation decision-making to recover from unplanned travel disruptions in the presence of ridesourcing options.

## 8. Conclusions

In this study we use a natural experiment design to identify the main determinants of ridesourcing substitution as an adaptive response strategy during unplanned transit disruptions. Our findings reveal that spikes in ridesourcing demand are strongly tied to peak-hour and weekday travel, suggesting a city-wide effect of mandatory travel. However, the relationship between transit disruptions and ridesourcing substitution behavior is also influenced by the temporal and spatial context in which the disruption occurs. Peak-hour disruption shifts to ridesourcing are positively correlated with the percentage of White residents in the surrounding community area, which suggests potential accessibility inequity in terms of options for mobility adaptation (i.e., resilience). That is, riders in areas with higher shares of non-White residents are less likely to turn to ridesourcing for mandatory travel during transit disruptions. This suggests a greater vulnerability of racial minorities to the negative impacts of transportation disruptions. Inequitable access to multiple (redundant) and robust mobility options needs to be addressed by transit agencies and transportation policymakers. In this paper we outline and discuss two policy perspectives arising from these findings, namely public-private partnerships and information campaigns, and a three-level framework for agencies to think more broadly about mobility justice initiatives.

**Acknowledgements**

This research was supported in part by funding from the National Defense Science and Engineering Graduate (NDSEG) fellowship provided to Borowski. Stathopoulos was partially supported by the US National Science Foundation (NSF) Career grant No. 1847537.

*Borowski et al*. 24Sun, H., Wu, J., Wu, L., Yan, X., & Gao, Z. (2016). Estimating the influence of common disruptions on urban rail transit networks. *Transportation Research Part A: Policy and Practice*, 94, 62-75.
U.S. Census Bureau. (2017). 2017 American Community Survey 5-Year Estimates, *American FactFinder*. Available at: http://factfinder.census.gov
van Exel, N. J. A., & Rietveld, P. (2001). Public transport strikes and traveller behaviour. *Transport Policy*, 8(4), 237-246.
Wampold, B. E., & Serlin, R. C. (2000). The consequence of ignoring a nested factor on measures of effect size in analysis of variance. *Psychological Methods, 5*(4), 425–433. https://doi.org/10.1037/1082-989X.5.4.425
Younes, H., Nasri, A., Baiocchi, G., & Zhang, L. (2019). How transit service closures influence bikesharing demand; lessons learned from SafeTrack project in Washington, DC metropolitan area. *Journal of Transport Geography*, 76, 83-92.
Zhang, Y., & Zhang, Y. (2018). Examining the relationship between household vehicle ownership and ridesharing behaviors in the united states. *Sustainability*, 10(8), 2720.
Zhao, F., Chow, L. F., Li, M. T., Ubaka, I., & Gan, A. (2003). Forecasting transit walk accessibility: Regression model alternative to buffer method. *Transportation Research Record*, 1835(1), 34-41.
Zhu, S., Masud, H., Xiong, C., Yang, Z., Pan, Y., & Zhang, L. (2017). Travel Behavior Reactions to Transit Service Disruptions: Study of Metro SafeTrack Projects in Washington, DC. *Transportation Research Record*, 2649(1), 79-88.